\documentclass[%
prst,
rsi,%
 amsmath,amssymb,
 reprint,%
]{revtex4-1}

\usepackage{subfig}
\usepackage{dcolumn}
\usepackage{bm}
\usepackage{graphicx}

\begin{document}

\preprint{AIP/123-QED}

\title[Sample title]{Temporal diagnostics of femtosecond electron bunches with complex structures \\using sparsity-based algorithm}

\author{Q. Q. Su}
\author{J. F. Hua}%
\email{jfhua@tsinghua.edu.cn}

 \author{Z. Nie}%
 \author{Y. Ma}%
 \author{S. Liu}
 \author{Y. F. ZHENG}
 \author{C. -H. Pai}
 \author{W. Lu}

\affiliation{
Department of Engineering Physics, Tsinghua University, Beijing 100084, China
}%

\date{\today}

\begin{abstract}
Femtosecond electron bunches with complex temporal structures play a crucial role in THz generation, free-electron lasers and plasma wakefield accelerators. The ultrashort electron pulse duration can be reconstructed from the coherent transition radiation (CTR) spectrum based on prior knowledge. A weighted greedy sparse phase retrieval (WGESPAR) algorithm is developed to reduce the ambiguities for reconstructing the distribution of the beam current. This algorithm achieves better performance than iterative algorithms, especially for truncated noisy spectra of multibunch structures. Based on the WGESPAR algorithm, the complex temporal structures of femtosecond electron bunches generated from laser wakefield accelerators can be successfully reconstructed.
\end{abstract}

\pacs{Valid PACS appear here}
\maketitle

\section{\label{sec:level1}Introduction}
Laser plasma accelerators\cite{lundh2011few} and bunch compression techniques\cite{PhysRevLett.111.184801} have a proven capability for the generation of few-femtosecond(fs) electron beams for plasma wakefield accelerators, ultrafast electron microscopy, compact ultrafast x-ray sources such as free-electron lasers and Thomson scattering sources\cite{PhysRevLett.116.184801,PhysRevLett.98.144801,ihee2001direct,sciaini2009electronic,PhysRevLett.114.195003,PhysRevLett.119.154801}. The temporal diagnostics of few-femtosecond electron bunch profiles hence have become a crucial issue in enhancing the performance of these facilities.


Methods of picosecond and sub-picosecond electron bunch length measurement, such as deflecting cavities, electro-optic techniques and frequency-domain methods\cite{berden2007benchmarking,yan2000subpicosecond,uesaka1998precise,PhysRevLett.93.114802}, have been well developed for conventional accelerators.  However, the temporal characteristics of few-femtosecond electron bunches cannot directly utilize these methods due to their limited resolution. Because the amplitude of the transverse deflecting voltage is one of the critical constraining factors, high powered lasers have been introduced for the desired sub-fs resolution\cite{dornmair2016plasma,zhang2016temporal,kotaki2015direct,zhang_du_tang_ding_huang_2017}. For electro-optic techniques, the resolution is strongly influenced by the crystal response function\cite{casalbuoni2005numerical}, and the achievable temporal resolution has been demonstrated down to a few tens of femtoseconds. Frequency-domain methods have the potential to achieve sub-femtosecond resolution by scanning the autocorrelation curve of coherent transition radiation (CTR), diffraction radiation or coherent synchrotron
radiation\cite{PhysRevLett.73.967,PhysRevE.63.056501,sei2017measurement}.

To measure the duration of few-femtosecond electron bunches, the method of CTR has frequently been adopted. CTR is generated when electron bunches propagate through inhomogeneous media, and the longitudinal bunch profile can be reconstructed from the attained CTR spectrum by  Kramer-Kronig relations\cite{lai1997using} or the fitting of Gaussian shape bunches \cite{lundh2011few,islam2015near,PhysRevSTAB.18.032802,maxwell2013coherent}. Conventionally, Kramer-Kronig relations work well in accelerator physics, but extrapolation to the unknown frequency ranges based on assumptions on the bunch shape is often required, which brings unavoidable uncertainty for complex temporal structures\cite{pelliccia2014two}. The Bubblewrap algorithm has been developed for the temporal characterization of femtosecond electron bunches, and a two-bunch structure can be reconstructed\cite{bajlekov2013longitudinal,PhysRevSTAB.18.121302}. However, the retrieved phase would be sensitive to noise and have ambiguities, which consequently lead to longer bunch lengths or even inaccurate bunch shapes. Recently, a greedy sparse phase retrieval (GESPAR) algorithm has been applied in coherent diffraction imaging and achieves good performance with  prior knowledge\cite{sidorenko_kfir_shechtman_fleischer_eldar_segev_cohen_2015}. The GESPAR algorithm can be heuristically applied in bunch duration reconstruction from the CTR spectrum.

In this paper, a WGESPAR algorithm is developed to reconstruct the temporal profiles of few-femtosecond electron bunches with complex structures. In Sec. \ref{sec:theory}, we describe the theory of CTR diagnostics and conduct a Monte Carlo simulation for CTR generated by ultrashort electron bunches passing through a radiator. In Sec. \ref{sec:algorithm}, a WGESPAR algorithm is developed to be specifically applied in the  reconstruction of complex longitudinal bunch structures from the CTR spectrum; the algorithm's performance  is compared with the Gerchberg-Saxton (GS) algorithm. The effectiveness of the WGESPAR algorithm is confirmed by the synthetic simulation of OSIRIS Particle-in-Cell (PIC) code and Monte Carlo code.

\section{\label{sec:theory}Theory and Monte Carlo simulation of CTR spectrum}
Transition radiation is emitted when charged particles pass through inhomogeneous media in the general case\cite{ginzburg1982transition}. In the transverse dimensionless case, the CTR energy spectrum density $\frac{dI}{d\omega d\Omega}$ observed in the direction of $\theta$ is given by

\begin{equation}\label{eq:distribution}
\frac{dI}{d\omega d\Omega} =N^2 \left| \rho(\omega)\right|^2 \frac{dI_e}{d\omega d\Omega}
\end{equation}

\begin{figure}[b] 
  \centering
  \includegraphics[width = 6cm]{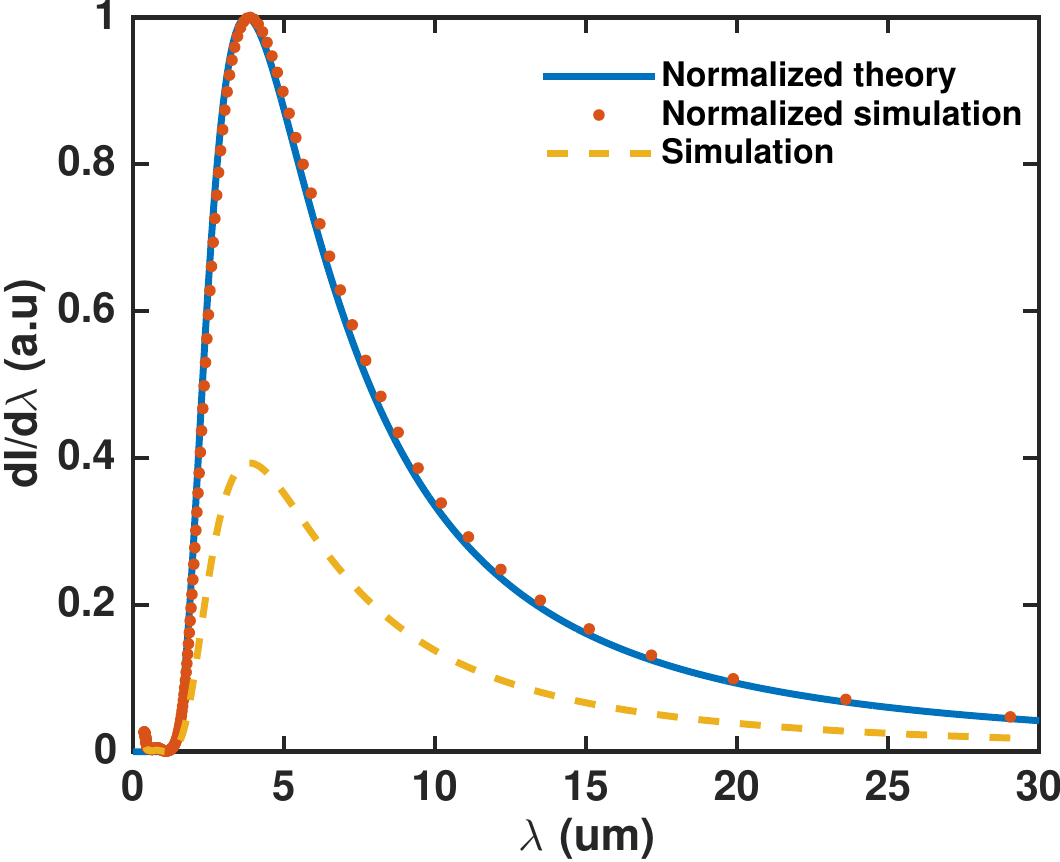}
\caption{\label{fig:monte_ex}\small{Normalized CTR spectrum calculated from theory (blue line) and Monte Carlo simulation (red dot). The original spectrum simulated by Monte Carlo code is divided by the peak of theoretical spectrum (yellow dashed line) as an evaluation of the energy loss due to the limited collection efficiency.}  }
\end{figure}

Here, $N$ denotes the total number of electrons and $\rho(\omega)$ is the form factor calculated by the Fourier transform of the electron density profile. Specifically, the form factor of a Gaussian bunch is $e^ {- \frac{\sigma_t^2 \omega^2}{2} }$, where $\sigma_t$ is the electron bunch duration. $\frac{dI_e}{d\omega d\Omega}$ means the spatial
energy distribution of the CTR generated by a single electron\cite{ginzburgorigin}:
\begin{equation}\label{eq:magfield}
\frac{dI_e}{d\omega d\Omega} = \frac{e^2}{2 \pi^2 c} ( \frac{\vec{\beta}  \times \hat{n} } {1 - \hat{n} \cdot \vec{\beta}}  -  \frac{\vec{\beta}^\prime  \times \hat{n} } {1 - \hat{n} \cdot \vec{\beta}^\prime})
\end{equation}

This is called the Ginzburg-Frank formula. Here, $\vec{\beta} $ and $\vec{\beta^\prime}$ denote the normalized velocity of the electrons and their image charges, respectively, and $\hat{n}$ is the unit vector pointed toward the observer. $\frac{dI_e}{d\omega d\Omega}$ is not correlated with the optical frequency; hence, the longitudinal profile is only related to the form factor $\rho(\omega)$.

\begin{figure}[t] 
  \centering
  \includegraphics[width = 7cm]{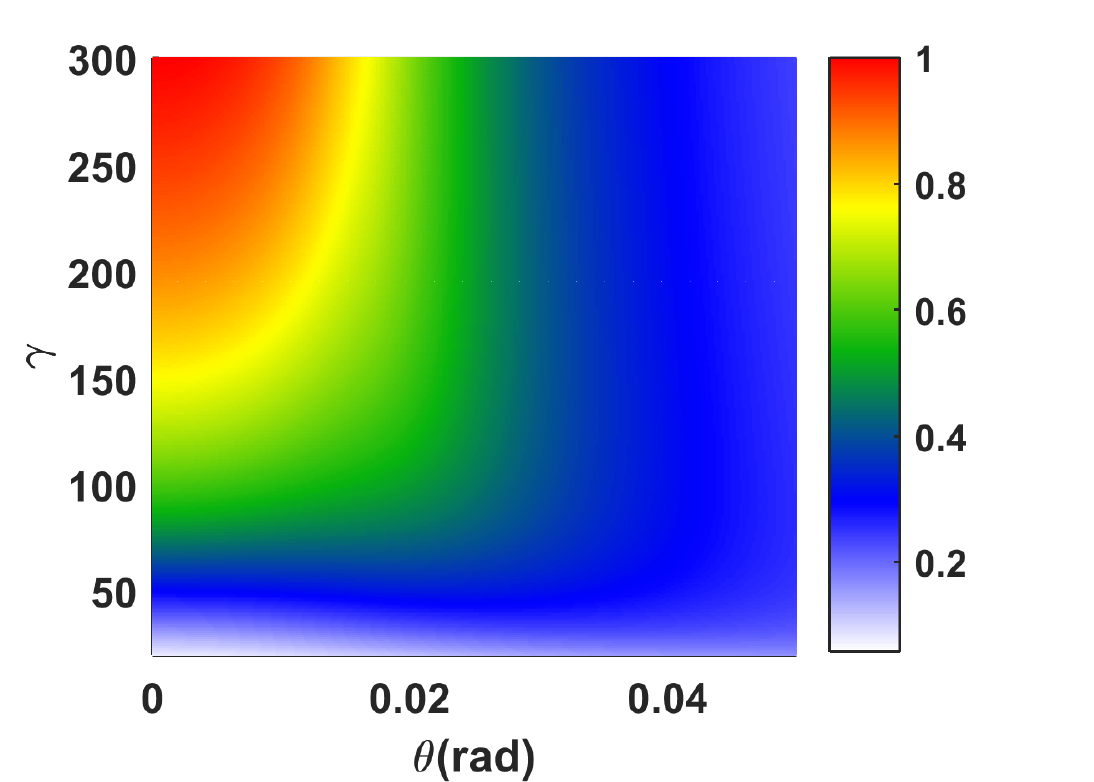}
\caption{\label{fig:proportion}\small{Collection efficiency considering the polar angle and the energy of the electrons. The CTR is generated by electrons whose energy varies from 20 to 150 MeV and whose polar angle between their velocity and the beamline is in the range of $0$-$0.05$ rad. } }
\end{figure}
Since the form factor of CTR is dependent on the distribution of electron bunches, the forward CTR radiation will inevitably be altered as the electron bunches pass through the radiator. Here, we use Al foil as the radiator. We  conduct a series of Monte Carlo simulations on Geant4\cite{lai1997using,agostinelli2003geant4} to estimate the actual CTR spectrum considering with the effect of the Al foil and bunch transverse distribution. In the simulation, CTR can be precisely calculated by transporting each electron separately through an Al foil and subsequently superposing the field of all electrons on the collection plane. Here, the collection plane is set to be 646 mm from the radiator, and the collection angle is 15 mrad. In Fig. \ref{fig:monte_ex}, 100 MeV electrons are emitted randomly from a point source with a divergence of 2 mrad (rms) and a duration of 2 fs (rms). After passing through a 76 um Al foil located 1 cm from the electron source, the arrival time, position and momentum of each electron on the rear surface of the foil are recorded. From the calculated CTR spectrum, we can find that the spatial distribution of electron bunches is nearly invariable, and thus, the CTR spectrum shape is close to the optimal case (Fig. \ref{fig:monte_ex}), validating this method for characterizing the longitudinal structure of electron beams  from the form factor. However, the simulated spectrum amplitude is nearly half the theoretical calculation under the influence of electron divergence and the limited collection angle. According to the simulation, a 10 pC electron bunch could generate more than 300 nJ CTR in the wavelength range of 1-9 um, which provides practical guidance for the implementation of beam temporal profile measurements such as the design of spectrometer layouts and the choice of mid-infrared detector.

From the Ginzburg-Frank formula, the opening angle of the CTR radiation cone can be defined as $\theta \sim\frac{1}{\gamma}$, where $\gamma$ stands for the Lorentz factor. Since the observation angle is limited by the spectrometer aperture, the collection efficiency will be influenced by the electron energy and the polar angle between the velocity direction and the beamline.  Considering the polar angle and electron energy, we quantify the collection efficiency as the ratio of the collected CTR energy in the preset collection angle to the total radiated CTR energy shown in Fig. \ref{fig:proportion}. To quantify the current profile reconstructed from the CTR spectrum, the effective charge can be regarded as the initial electron charge multiplied by the square root of the collection efficiency ratio. Combining the concept of effective charge with phase retrieval algorithms, the shape of the beam current successfully reconstructed from the obtained CTR spectrum should be the same as the current contributed by the total effective charge.

\section{\label{sec:algorithm}Algorithms for longitudinal bunch profile reconstruction}

The reconstruction of the beam current from the CTR spectrum can be regarded as one-dimensional phase retrieval problems, which have been studied for decades. The reconstruction process can be summarized as seeking a current profile $\rho(t)$, the Fourier transform $\hat{\rho}(\nu)= \int_{-\infty}^{+\infty} \rho(t) \exp{\left(-2\pi i \nu t\right)} dt$ of which has the same amplitude as the measured CTR spectrum $\left|\hat{g}(\nu)\right|$, and the missing phase information is expected to be retrieved from the CTR spectrum modulus.  For physical considerations, the function $\rho(t)$ must be nonnegative and constrained in a finite support.

Several algorithms, such as the Kramer-Kronig relations\cite{lai1997using} and iterative algorithms, including the GS algorithm  and  Hybrid Input-Output Algorithm (HIO)\cite{Fienup:82}, which have been applied in astronomy, X-ray diffraction imaging and other applications, have been developed to solve the phase retrieval problem. The emergent Bubblewrap algorithm is another iterative algorithm based on the GS and HIO algorithms. Iterative algorithms can reconstruct longitudinal bunch profiles with sufficient frequency domain information, but they may lead to longer bunch tails if low-frequency information is missed. Recently, it has been proved that one-dimensional phase retrieval problems still have intrinsic ``zero-flipping" ambiguities even with the assumption of finite support and non-negativity\cite{walther1963question,beinert2017enforcing}. With the development of compressed sensing in the past few years, sparsity-based algorithms\cite{candes_eldar_strohmer_voroninski_2015} have been widely applied in the field of signal processing because the reconstruction is robust even with highly noisy spectra. Accordingly, optimization-based algorithms, such as semi-definite programming, the GESPAR algorithm and the sparse Fienup algorithm \cite{shechtman2014gespar},  have been proposed to solve phase retrieval problems with a sparsity priori. The GESPAR algorithm has been recently adapted in coherent diffraction imaging and achieved good performance with sparsity as prior knowledge\cite{sidorenko_kfir_shechtman_fleischer_eldar_segev_cohen_2015}. We develop a weighted GESPAR (WGESPAR) algorithm for electron bunch profile reconstruction, and we compare its performance with the GS algorithm.

\subsection{One-dimensional phase retrieval algorithms}

%
The extensively used GS algorithm and HIO algorithm are error-reducing algorithms used to obtain a better approximation at each iteration. Suppose that the Fourier transform of the bunch profile $\rho(t)$ is given by  $\hat{\rho}(\nu) = \left|\hat{g}(\nu) \right| e^{-\phi(\nu)} $. Here, $\rho(t)$ is a real, non-negative signal with compact support, and $\left| \hat{g}(\nu) \right|$ is the square root of the band-limited spectrum measured in the experiment. A random function $\rho_0(t)$ can act as the initial beam current profile.  The k-th trial solution $\rho_k(t)$ is Fourier transformed, yielding $\left| \hat{\rho}_k(\nu) \right| e^{-\phi_k(\nu)}$. To satisfy the Fourier-domain constraints, a better estimation of $\hat{\rho}(\nu)$ is given by $\hat{\rho}^\prime _k(\nu) = \left| \hat{g}(\nu) \right| e^{-\phi_k(\nu)} $. Then, the resulting $\hat{\rho}^\prime_k(\nu)$ is inverse Fourier transformed, yielding the function $\rho^\prime_{k}(t)$. $\rho^\prime_{k}(t)$ is then modified to satisfy the time-domain constraints, yielding $\rho_{k+1}(t)$, which is regarded as the input value for the next iteration. Normally, a relaxation factor $r$ is introduced to improve performance. The relaxation iterative format is called an HIO algorithm. When $r = -1$, the HIO algorithm simplifies to the GS algorithm.

In iterative algorithms, such as GS and HIO, overfitting usually occurs, leading to unphysical results when some information is missing and the spectrum data are highly noisy. From a different perspective, the GESPAR algorithm considers the phase retrieval as a nonlinear least-squares optimization problem, therein minimizing the error $ E = \left \| \bm{\hat{\rho}}^2 - \left|\bm{\hat{g}}\right|^2 \right \|$ between the square of the discrete Fourier transform of the reconstructed signals $\bm{\hat{\rho}}^2$ and the measured spectrum $\left|\bm{\hat{g}}\right|^2$ in a confined set of signals. Here, both $\bm{\hat{\rho}}$ and $\bm{\hat{g}}$ are vectors.  Substituting $\bm{\hat{\rho}} = F\bm{\rho}$ as the discrete fourier transform(DFT) of the time-domain signal $\bm{\rho}$ by multiplying it with the DFT matrix $F$, the objective optimization function can be written as $ E = \sum^N _{i = 1}(\bm{\rho}^T F_i^T F_i \bm{\rho} - \hat{g}_i^2) $. Here $F_i$ is the $i$th row of the DFT matrix $F$. The non-negative signal vector $\bm{\rho}$ can be approximated by a projection on a specific base set, for example, a base set of Gaussian bunches. The approximated signal takes the form $ \bm{\rho} = D\bm{x}$, where $D$ is the dictionary of the preset base set and $\bm{x}$ means the projection of the signal $\bm{\rho}$ onto that base. We make an assumption that the electron current profile can be approximated by a linear combination of less than $s$ bases defined by the dictionary $D$. With those assumptions, the approximation of the signal $\bm{\rho}$ can be solved uniquely by minimizing the error function $E$.

\begin{figure}[t] 
  \centering
  \includegraphics[width = 7cm,height = 8cm]{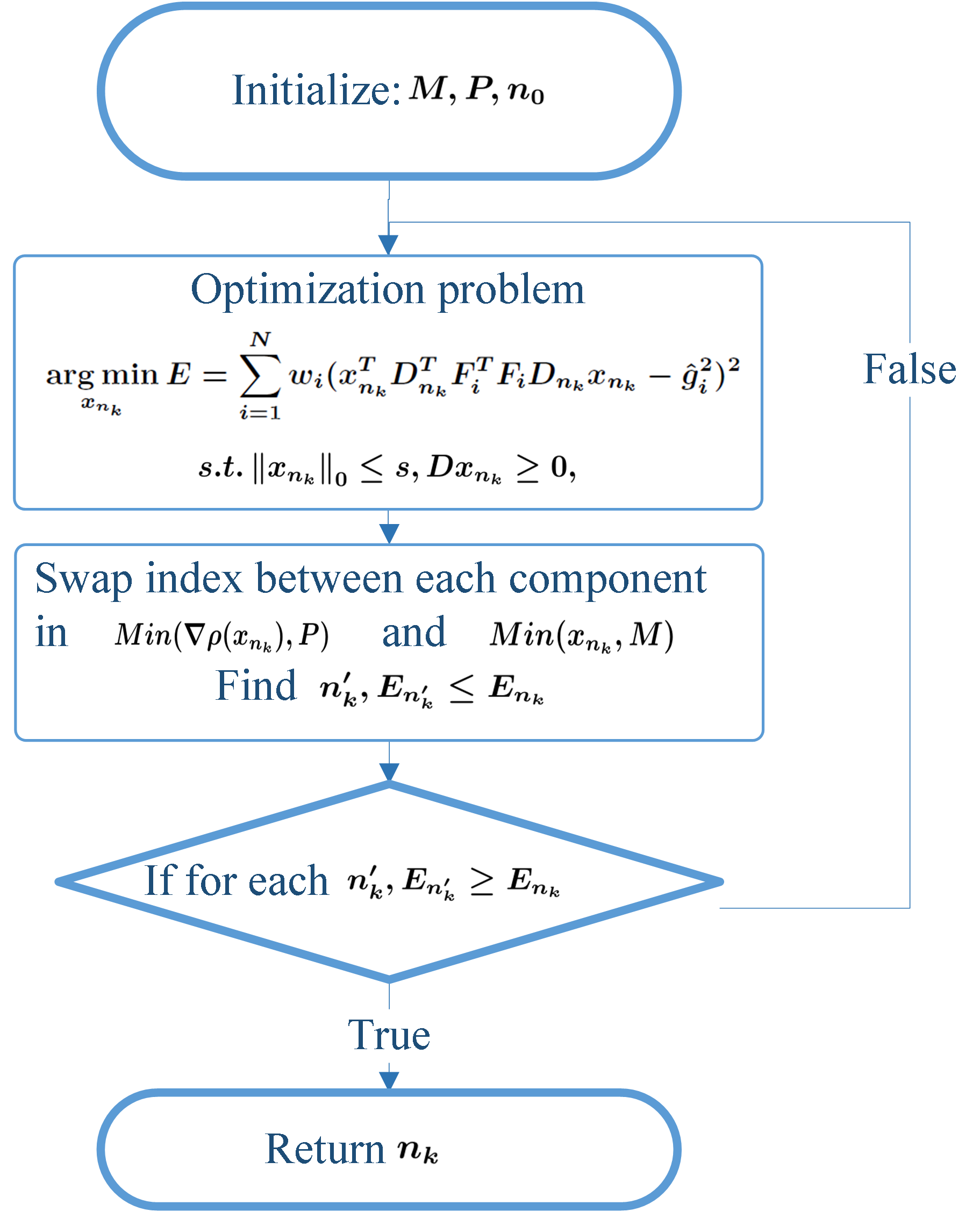}
  \caption{\label{fig:GESPAR}\small{ Flow chart of the WGESPAR algorithm. $Min(A,B)$ denotes the index of the $B$ components of $A$ with the lowest value.}}
\end{figure}

For  practical spectrum measurements, the noise introduced by the detector should be considered. According to the relation $d\omega = -\frac{2 \pi c}{\lambda^2}d\lambda$, the spectrum data acquired in the experiment will be characterized by different noise levels, which will influence the reconstruction results if all the spectrum data are given the same weight. To solve this problem, the optimization function is rewritten as the weighted square of the error between the reconstructed spectrum and the approximated real spectrum $ E = \sum^N _{i = 1} w_i(\bm{\rho}^T F_i^T F_i \bm{\rho} - {\hat{g}_i}^2)^2 $, where $w_i$ is the normalized weight introduced as an evaluation of the importance of each data point. Here, $w_i = \frac{1}{\sigma_i^2 (\sum_{i = 1}^n \frac{1}{\sigma_i^2} )}$, where $\sigma_i$ stands for the standard variance of spectrum data $i$. Finally, we obtain the mathematical formulation of the phase retrieval problem subject to the non-negativity, finite support and sparsity constraints:

 \begin{align}
\mathop{\arg\min}_x E &= \sum^N _{i = 1}w_i(\bm{x}^T D ^T F_i^T F_i D \bm{x} - {\hat{g}_i}^2)^2,\nonumber \\
w_i &= \frac{1}{\sigma_i^2 (\sum_{i = 1}^n \frac{1}{\sigma_i^2} )}\\
s.t.   \left \| \bm{x} \right \|_0 &\leq s, \nonumber\\
       D\bm{x} &\geq 0, \nonumber
\end{align}

Here, $\left \| \bm{x} \right \|_0$ stands for the number of non-zero elements of $\bm{x}$. Note that the dictionary is chosen in the support so that the support constraint can be automatically satisfied. For example, if the dictionary is chosen as a matrix based on Gaussian pulses with finite support, the reconstructed signal can only be the linear combination of those Gaussian signals such that it also satisfies the support priority.

\begin{figure}[b]
    \centering
    \subfloat{
    \begin{minipage}[b]{4cm}
        \includegraphics[width=3.7cm]{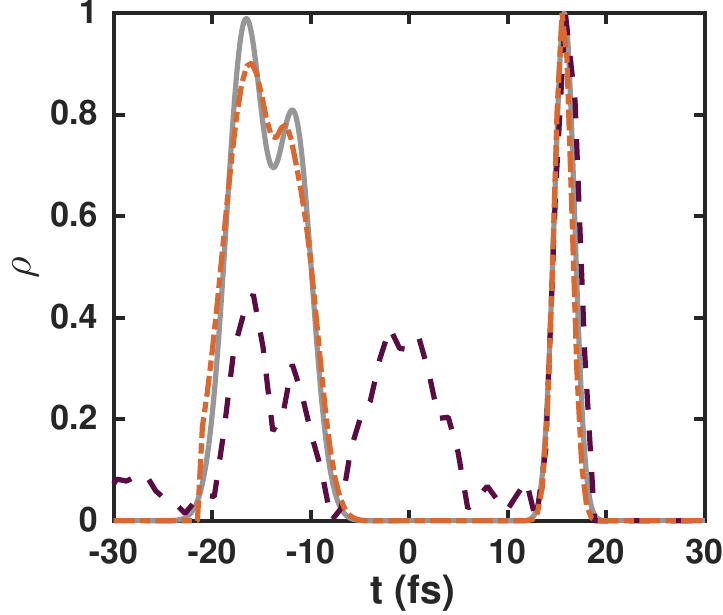}
        \caption*{(a)}
        \vspace{0.1cm}
        \includegraphics[width=3.7cm]{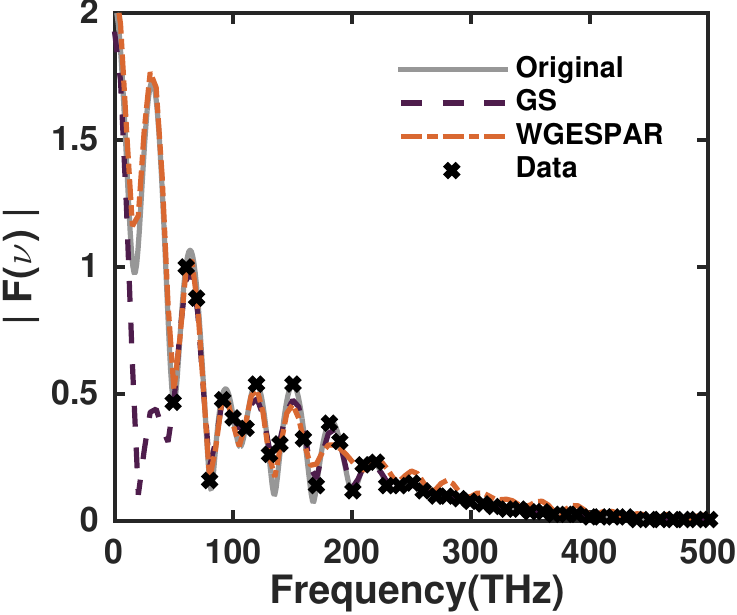}
        \caption*{(c)}
    \end{minipage}
    }
    \subfloat{
        \begin{minipage}[b]{4cm}
        \includegraphics[width=3.7cm]{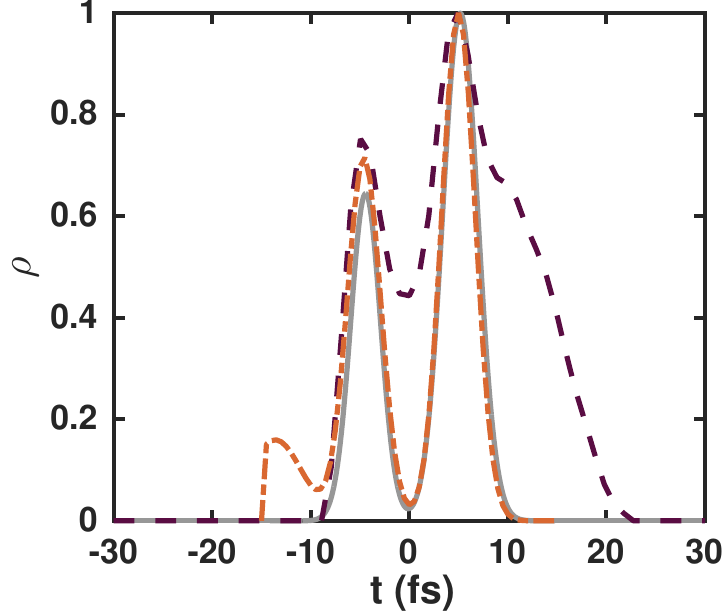}
        \caption*{(b)}
        \vspace{0.1cm}
        \includegraphics[width=3.7cm]{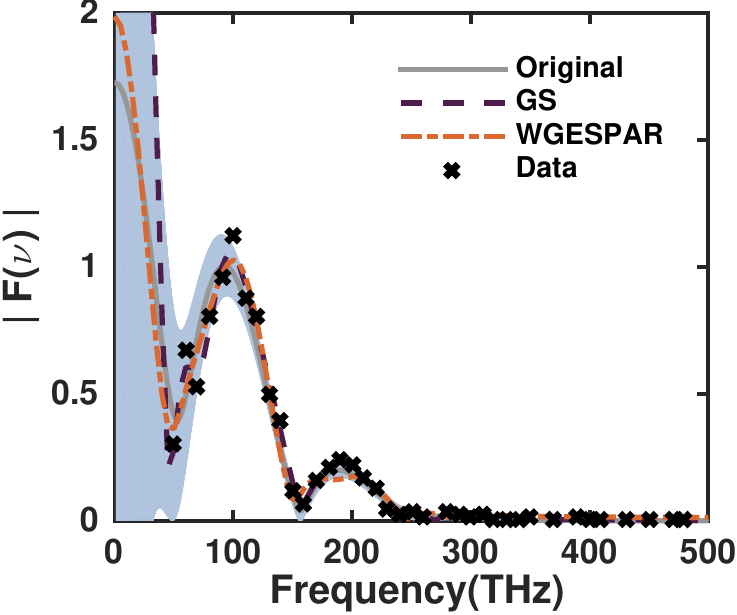}
        \caption*{(d)}
        \end{minipage}
    }
    \caption{\label{fig:compare}\small{The comparison of the WGESPAR and GS algorithms for longitudinal bunch profile reconstruction with large support (a,c) or a noisy spectrum (b,d). The beam currents (a,b) and the spectra (c,d) are both provided for the original signals (gray line) and the results reconstructed by the WGESPAR algorithm (red dashed line) and GS algorithm (purple dashed line). The cross spots are spectrum data $\left|F(\nu) \right|$ ``measured" virtually in the simulation. The blue shading represents the uncertainty of the measured CTR spectrum when 10\% noise is added to the virtual detector. }}
\end{figure}

The flow chart of the WGESPAR algorithm is shown in Fig. \ref{fig:GESPAR}, which is different from its original form\cite{bajlekov2013longitudinal} considering the non-negative condition and the weighted optimization function. Let $\bm{n}$ be the index array of the non-zero elements of the vector $\bm{x}$, where $\bm{x_{n}}$ means the non-zero element vector of $\bm{x}$  such that the zero norm of $\bm{x}$ is equal to the length of the index vector $\bm{n}$: $\left \| \bm{x} \right \|_0 = \left|\bm{n}\right|$. The index array has a random initialization value $\bm{n_0}$ with a length of $s$. At the $k$-th iteration, the optimization problem $\mathop{\arg\min}\limits_{\bm{x_{\bm{n}_k}}} E = \sum^N _{i = 1}  w_i(\bm{x_{n_k}} ^T D_{\bm{n}_k}^T F_i^T F_i D_{\bm{n}_k} \bm{x_{n_k}} - {\hat{g}_i}^2)^2$ is solved by the non-linear least-square trust-region algorithm\cite{subramanian1993gauss} and yields $\bm{x_{n_k}}$ and $E_{\bm{n_k}}$ with index array $\bm{n}_k$. Then, we swap the index of $M$ components of $\bm{x_{n_k}}$ with the smallest absolute value with the $P$ negative maximums of $\nabla E(x)$. After this swap, the index array becomes $\bm{n}^\prime_k$, and $E_{\bm{n}^\prime_k}$ is yielded by solving the optimization function again. If the optimization function value $E_{\bm{n}^\prime_k}$ is smaller than $E_{\bm{n}_k}$, the index vector $\bm{n}^\prime_k$ is reserved as the non-zero element index of the next loop $\bm{n}_{k+1}$ until the value $E_{\bm{n}^\prime_k}$ of each swap is higher than the original $E_{\bm{n}_k}$, and we accept $D\bm{x_{n_k}}$ as the reconstructed beam current.

\subsection{Comparison of WGESPAR and GS algorithms \label{sec:compare}}

To compare the WGESPAR algorithm with the GS algorithm, we generate pulses with random positions and durations in a finite support. The spectra of those test signals are sampled with a frequency step of 10 THz deduced from the resolution of a commercial mid-infrared detector. We apply the GS and WGESPAR algorithms to rebuild the phase of the truncated spectra within the range of 50 THz-500 THz. In the GS algorithm, we do not apply restrictions in the frequency range below 50 THz. The base of the WGESPAR algorithm is composed of Gaussian pulses. There are two ranges of pulse widths in that base, one from 1 fs to 4 fs with 0.5 fs increments and  one from 0.5 fs to 1 fs with 0.1 fs increments. The positions of all Gaussian pulses are mapped uniformly on the support interval, and the position interval is set to be 0.5 fs. The dictionary is generated corresponding to the Gaussian base. During the reconstruction, the WGESPAR algorithm is run  several times until the error is less than a preset threshold or the run times are larger than a preset number $N$. We set $M = 2, P = 30, N = 5$. $M$ and $P$ depend on the sparsity and complexity of the selected base.

For the GS algorithm,  the retrieved pulse length has been mentioned to be elongated because of a lack of information at longer wavelengths\cite{bajlekov2013longitudinal}. From Fig. \ref{fig:compare}, we can find the same phenomenon. In contrast, with the assumption of bunch series whose pulse width of every single bunch is less than 5 fs, bunch trains with a support length of 40 fs can be well reconstructed by WGESPAR even without frequency information below 50 THz.

\begin{figure}[bt]
    \centering
    \subfloat{
    \begin{minipage}[b]{4cm}
        \includegraphics[width=3.7cm]{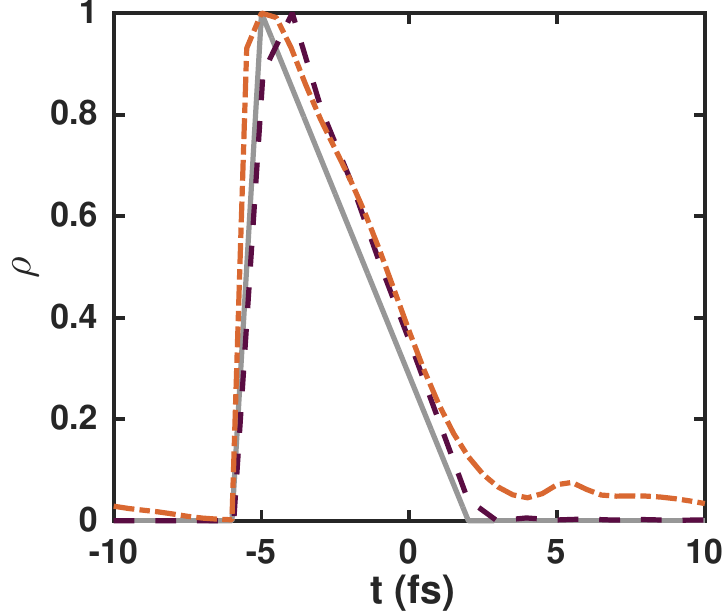}
        \caption*{(a)}
        \vspace{0.1cm}
        \includegraphics[width=3.7cm]{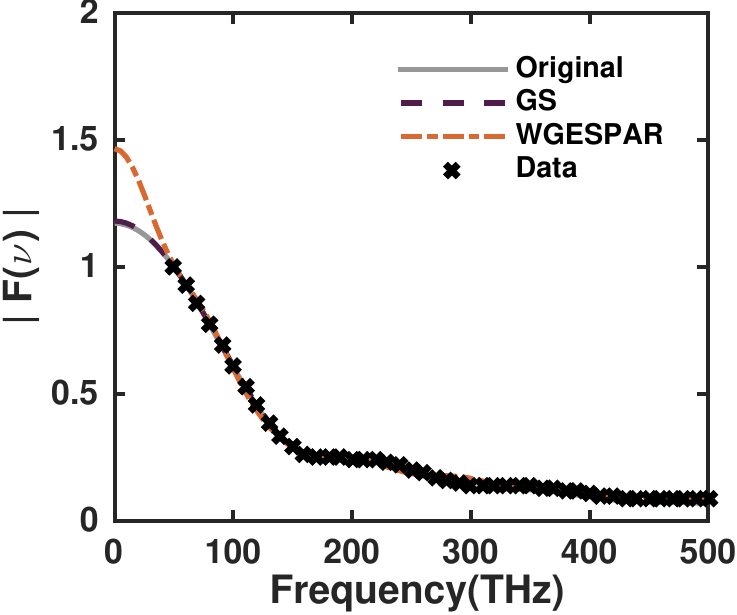}
        \caption*{\small{(c)}}
    \end{minipage}
    }
    \subfloat{
        \begin{minipage}[b]{4cm}
        \includegraphics[width=3.7cm]{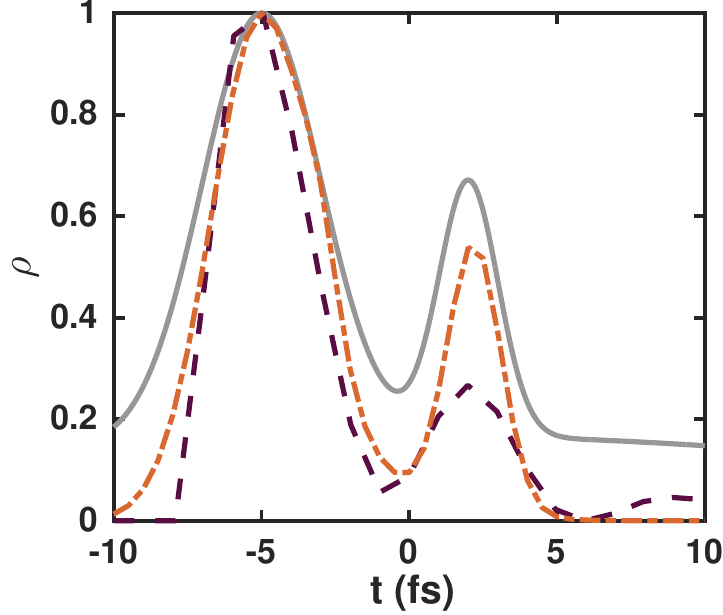}
        \caption*{(b)}
        \vspace{0.1cm}
        \includegraphics[width=3.7cm]{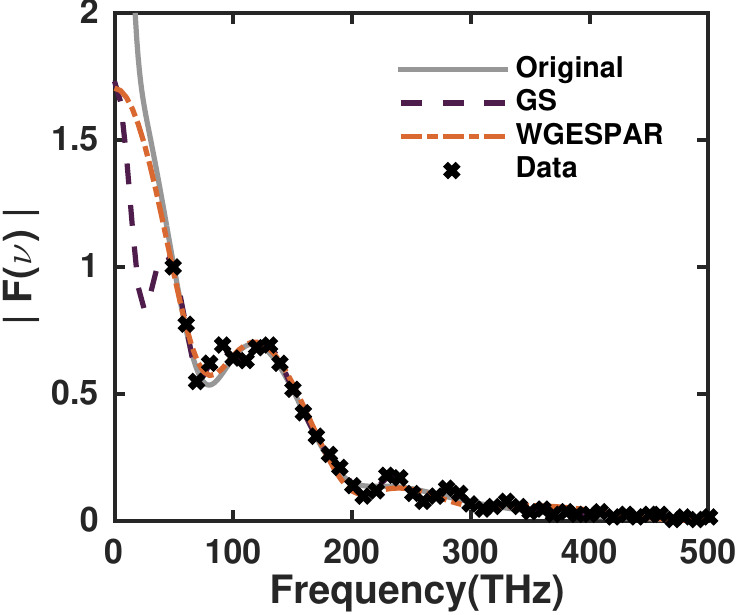}
        \caption*{(d)}
        \end{minipage}
    }
    \caption{\label{fig:special shape}\small{Reconstruction for a triangular bunch structure (a,c) and short bunches with a long current tail (b,d). The beam currents (a,b) and the spectra (c,d) are both provided for the original signals (gray line) and the results reconstructed by the WGESPAR algorithm (red dashed line) and the GS algorithm (purple dashed line). The cross spots are spectrum data $\left|F(\nu) \right|$ ``measured" virtually in the simulation. }}
\end{figure}

In the experiment, the influence of noise should be seriously considered for bunch profile reconstruction. Typical MIR pyroelectric cameras usually have a noise level of a few nJ, which is only two orders less than the expected CTR energy. To test the stability of these two algorithms, $10\%$ gaussian noise is added to the virtual ``detector" pixel. When the frequency goes to zero, the noise of the spectrum goes to infinity because of the relationship $d\omega = -\frac{2 \pi c}{\lambda^2}d\lambda$. Fig. \ref{fig:compare}(b,d) shows the temporal distribution rebuilt from the noisy spectrum. We can find that the GS algorithm is more sensitive to noise and fails to reconstruct the signal, even though it can work for the ideal case, whereas the WGESPAR algorithm tends to show a better tolerance to noise. This is because iterative algorithms conform to all noisy data of the spectrum, whereas optimization methods only find the signal in the preset group whose spectra are closest to the measured spectra.

Electron beams with specific current profiles have special applications in plasma wakefield accelerators and THz generators\cite{PhysRevLett.116.184801,PhysRevLett.98.144801}. In Fig. \ref{fig:special shape}, WGESPAR shows good performance for special bunch profile reconstruction such as triangular and multi-bunch spikes superposed on long bunch backgrounds. The reconstructed beam current maintains main structural features such as the slope of the bunch and the bunch train intervals. If other priories (such as an approximation of the bunch shape) are provided, the reconstruction result will be improved.

\subsection{\label{sec:simulation}Beam current reconstruction for combined simulation of PIC and Monte Carlo}

\begin{figure*}[t] 
  \centering
  \begin{minipage}{18cm}
    \begin{minipage}{5cm}
  \includegraphics[width = 5cm,height = 3cm]{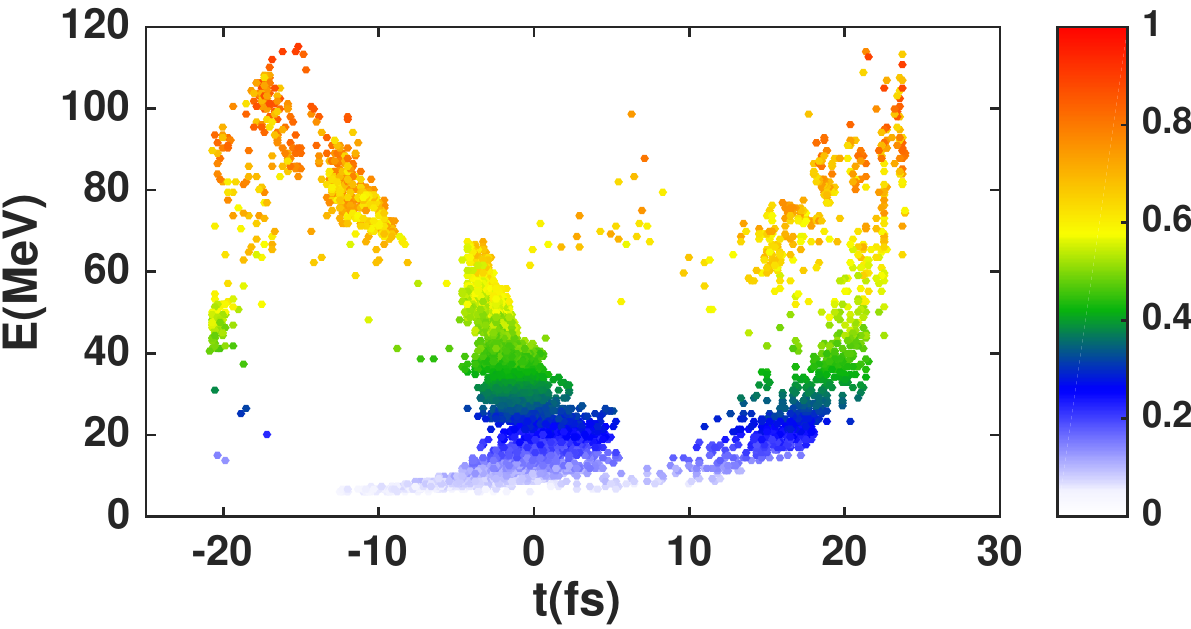}
  \caption*{(a)}
    \end{minipage}
    \begin{minipage}{5cm}
  \includegraphics[width = 5cm,height = 3cm]{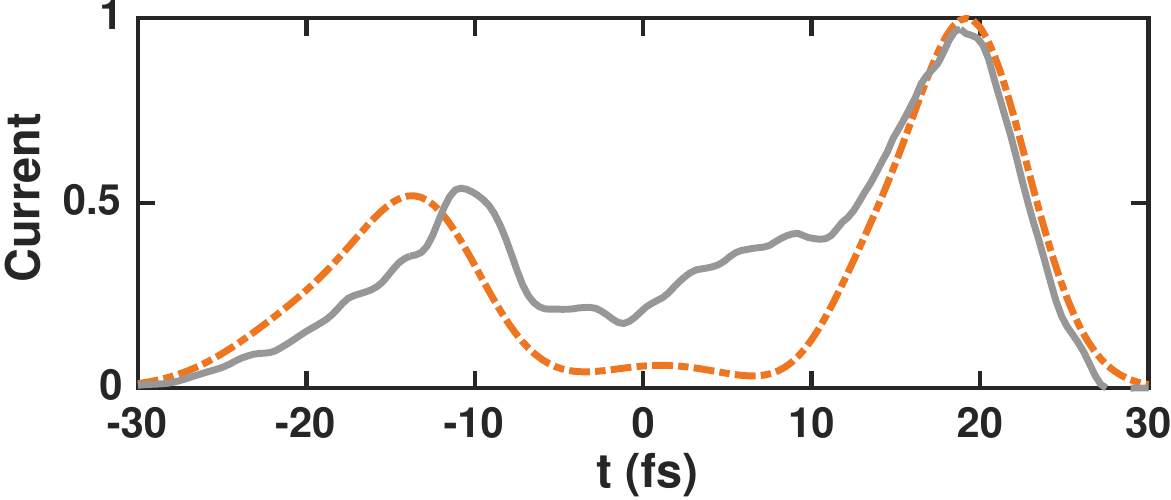}
  \caption*{(b)}
    \end{minipage}
    \begin{minipage}{5cm}
  \includegraphics[width = 5cm,height = 3cm]{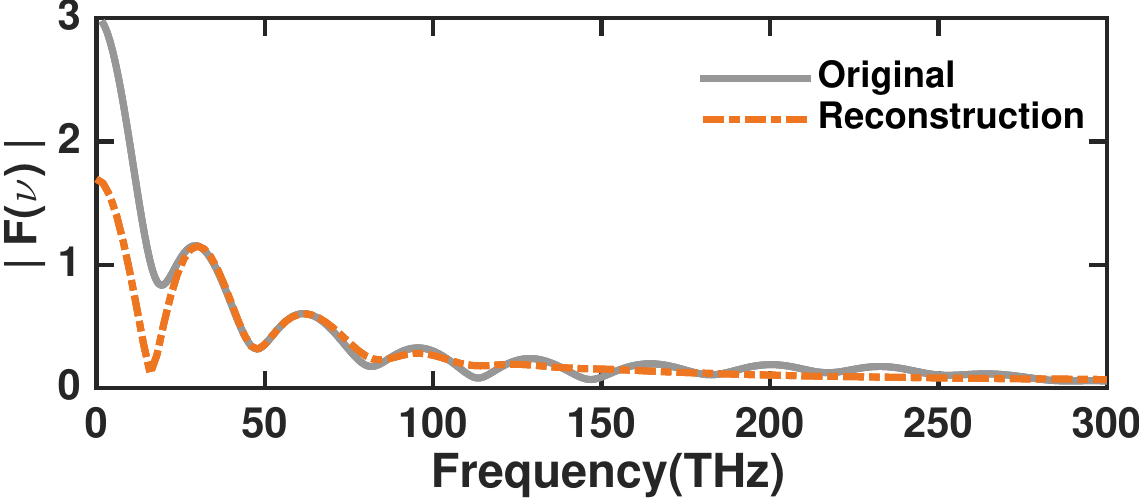}
  \caption*{(c)}
    \end{minipage}
  \end{minipage}
  \caption{\label{fig:simu_reconstruction}\small{The longitudinal phase space of a typical simulated electron bunch(a) and the comparison of the original (gray line) and reconstructed beam currents (red dashed line). Each electron is indicated with different colors based on their different effective charge. The electron current profile(b) is reconstructed from the CTR spectrum(c) simulated by OSIRIS and Geant4. }}
\end{figure*}

In the previous section, we  demonstrate that the WGESPAR algorithm can effectively reconstruct complex current structures by reducing the number of feasible solutions. To test the WGESPAR algorithm for the reconstruction of real beam currents, we have performed 2D PIC simulations for downramp injected electron beams using OSIRIS. A 35 fs (rms), 800 nm laser is focused at the front edge of a plasma using a spot with a diameter of 10 um (rms), and the peak plasma density is set as $1.74 \times 10^{19}$ cm$^{-3}$, with a sharp transition at 100 um. Strong downramp injection of electrons with a two-pulse current structure have been observed in the simulation. The injected quasi-monoenergetic electron bunch can be accelerated to 100 MeV with an energy spread of 40\%(Fig. \ref{fig:simu_reconstruction}(a)). Using the accelerated electron bunches as the radiation source, the process of CTR generation is simulated by the Monte Carlo method discussed in Sec. \ref{sec:theory}. Due to the limited collection aperture for CTR collection in experiment,  electrons with energy higher than 20 MeV are considered as the CTR source. The longitudinal electron bunch profiles are reconstructed from spectra simulated by the Monte Carlo code with a sampling rate of 5 THz from 35 THz to 350 THz.

We use the WGESPAR algorithm for phase retrieval with a support of $ \left[-30,30\right]$ fs and the non-negativity constraint.  A Gaussian base with pulse width varying from 1 fs to 4 fs, with an increment of 0.5 fs, is selected. We set the WGESPAR algorithm parameters as follows: M = 2, P = 30 and N = 5.  The sparsity parameter varies from 2 to 10, and the result with the lowest sparsity whose spectrum deviation from the measured spectrum is within a preset threshold is chosen. 
Fig. \ref{fig:simu_reconstruction} shows a typical reconstruction result.

We have previously discussed the dependence of the CTR spectrum on the electron energy and the polar angle. The effective charge is introduced by taking these two factors into account. In the sense of effective charge, the WGESPAR algorithm can successfully reconstruct the main features of two peak beam currents (Fig. \ref{fig:simu_reconstruction}), where the triangular shape of the sub-bunch can also be clearly recognized. The relative deviations of the current amplitude, beam intervals and bunch length of each sub-bunch are less than 5\%. A plateau after the main bunch is missing because of the lack of information under 35 THz.

\section{\label{sec:conclusion}Conclusions}
It has been proved that fs beam current profiles can be reconstructed using a CTR spectrum with sufficient bandwidth. However, the achieved spectra in the experiment are usually truncated due to the experimental layouts, which can lead to the inaccurate reconstruction of complex longitudinal bunch structures. An optimization-based algorithm, WGESPAR, has been developed for the temporal characterization of electron bunches even with truncated noisy spectra. We find that the WGESPAR algorithm exhibits better tolerance than iterative algorithms to noisy spectra with a sparsity assumption. Using the synthetic simulation with Particle-in-Cell and Monte Carlo codes, the WGESPAR algorithm can be successfully applied to measure ultrashort electron bunches with complex temporal structures generated in laser-plasma accelerators, thereby paving the way for  applications in ultrafast science with fs electron beams.

\section{\label{sec:conclusion}Acknowledgement}
We thank Dr Yingxin Wang for providing the Pyrocam III detector. This work was supported by NSFC Grant No.11425521, No. 11535006, No. 11375006, No. 11775125 and No. 11475101.

\nocite{*}
\bibliography{ref}

\end{document}